\documentclass[twocolumn,showpacs,preprintnumbers,amsmath,amssymb]{revtex4}
\usepackage{graphicx}

\begin{document}

\normalsize{Phone Number: ++86-10-8823-6167}

\normalsize{ Fax Number: ++86-10-8823-3085}

\normalsize{ E-mail: caolg@mail.ihep.ac.cn}

\vglue 1.0cm
\title{Isovector Giant Dipole Resonance of Stable Nuclei in a Consistent Relativistic Random Phase Approximation}
\author{Li-Gang Cao$^{1}$  Zhong-Yu Ma$^{2,3,4}$}

 \affiliation{${}^1$ Institute of High Energy Physics,Chinese Academy of Sciences, Beijing 100039}
 \affiliation{${}^2$ China Institute of Atomic Energy, Beijing 102413}
 \affiliation{${}^3$ Center of Theoretical Nuclear Physics, National
Laboratory of Heavy Ion Accelerator of Lanzhou,Lanzhou 730000}
 \affiliation{${}^4$ Institute of Theoretical Physics, Chinese Academy of
Sciences, Beijing 100080}
\date{\today}

\begin{abstract}
A fully consistent relativistic random phase approximation is
applied to study the systematic behavior of the isovector giant
dipole resonance of nuclei along the $\beta$-stability line in
order to test the effective Lagrangians recently developed.  The
centroid energies of response functions of the isovector giant
dipole resonance for stable nuclei are compared with the
corresponding experimental data and the good agreement is
obtained. It is found that the effective Lagrangian with an
appropriate nuclear symmetry energy, which can well describe the
ground state properties of nuclei, could also reproduce the
isovector giant dipole resonance of nuclei along the
$\beta$-stability line.
\end{abstract}


\maketitle

The investigation both experimentally and theoretically on various
modes of nuclear collective giant resonances has become one of
major research fields in the nuclear structure physics since the
isovector electric (non-spin flip) giant dipole resonance (IV GDR)
had been investigated firstly by Baldwin and Klailer\cite{Bal47}
at the end of 1940's. The isoscalar electric giant monopole
resonance (IS GMR) in heavy nuclei is a unique direct source of
experimental information on the compression modulus and the
equation of state. The electric giant dipole resonance (GDR) and
giant quadrupole resonance (GQR) are relevant to the symmetry
energy coefficient and effective mass of nucleon, respectively.
Much is already known about the electric giant resonance,
especially for the isoscalar GMR, GQR and isovector GDR in cold
nuclei, where exist extensive data sets on the excitation energy,
strength distribution and width. The systematic behavior of the
isovector GDR was reviewed by Berman and Fultz\cite{Ber75} in
1975. The latest one was given by Speth\cite{Spe90} in 1991. On
the other hand, many theoretical works have been done to
understand the physics mechanism of various collective vibrations.
Furthermore, by comparing with the experimental data, one can test
the microscopic theory and effective interactions which are used
to reproduce the experimental data. One of the effective methods
used to study properties of the giant resonance is the random
phase approximation (RPA).

In recent years, the relativistic mean field (RMF) theory with
non-linear meson self-interactions has achieved a great success in
describing bulk properties of nuclei, not only spherical but also
deformed nuclei and nuclei far from the $\beta$-stability
line\cite{Ring96}. In particular, the RMF theory has also been
applied in studying the dynamical processes of nuclei. The linear
response of a system to an external field can be calculated in the
relativistic RPA. Early investigations using the relativistic
RPA\cite {Kur85,Hui89,She89} were based on Walecka's linear
$\sigma $-$\omega $ models, which provides considerably larger
incompressibility\cite{Ser86}. Therefore, one could not expect to
obtain quantitative agreement with experimental data in those
early calculations. Recently the meson propagators with non-linear
self-interactions have been worked out numerically and the
relativistic RPA calculations with the non-linear terms were
performed \cite{Ma97,MTG97,GMT99}. A fully consistent relativistic
RPA has been established in the sense that the relativistic mean
field wave function of nucleus and particle-hole residual
interactions in the relativistic RPA are calculated from a same
effective Lagrangian\cite{Ma99,MGR01}. A consistent treatment of
relativistic RPA within the RMF approximation requires the
configurations including not only the pairs formed from the
occupied Fermi states and unoccupied states but also the pairs
formed from the Dirac states and occupied Fermi states. It has
been formally proved \cite{RGM01} that the fully consistent
relativistic RPA is equivalent to the time dependent RMF (TDRMF)
at the small amplitude limit \cite{Vre99,Lal97}. The response
functions for the closed shell nuclei \cite{Ma99,MGR01} have been
calculated using the fully consistent relativistic RPA. The
isoscalar GMR, GQR and isovector GDR for some double magic
nuclei, such as $^{208}$Pb, $^{144}$%
Sm, $^{114}$Sn, $^{90}$Zr were performed with different effective
Lagrangian parameter sets NL3\cite{Lal97}, NL1\cite{Rei86},
NL-SH\cite{Sha93}, and TM1\cite{Sug94}. A good agreement with
experimental data is obtained\cite{MCG01}.

In this letter, we aim at the investigation on the systematic
behavior of the isovector GDR in $\beta$-stable nuclei by focusing
our attention to the centroid energies of response functions. By
evaluating various moments of the response functions, we extract
the centroid energies of response functions for each nucleus and
compare with the available experimental data and discuss the
systematic behavior. The method employed in our investigation is
the fully consistent relativistic RPA\cite{Ma97,MTG97,GMT99},
which is a relativistic extension of non-relativistic RPA for
studying microscopically nuclear dynamic excitations and giant
resonances. The motivation of this work is to test the effective
Lagrangians recently developed in a more wide case. We shall
investigate  the electric GDR relevant to the symmetry energy
coefficient.

The response function of a quantum system to an external field is
given by the imaginary part of the polarization operator,
\begin{equation}
R^{L}(Q,Q;{\bf k},{\bf k^{\prime }};E)=\frac 1\pi Im\Pi ^R(Q,Q;{\bf k},{\bf %
k^{\prime }};E)~,   \label{eq1}
\end{equation}
where Q is an external field operator. The relativistic RPA
polarization operator is obtained by solving the Bethe-Salpeter
equation,
\begin{eqnarray}
&&\Pi (Q,Q;{\bf k},{\bf k^{\prime }},E)=\Pi _0(Q,Q;{\bf k},{\bf
k^{\prime }}
,E)  \nonumber\\
&&-\sum_ig_i^2\int d^3k_1d^3k_2\Pi _0(Q,\Gamma ^i;{\bf k},{\bf k}_1,E) \nonumber\\
&&D_i({\bf k}_1,{\bf k}_2,E)\Pi (\Gamma _i,Q;{\bf k}_2,{\bf
k^{\prime }},E)~, \label{eq2}
\end{eqnarray}
where the i runs all mesons. In order to obtain the centroid
energies of the response functions,  we first calculate various
moments of the response function in a given energy interval,
\begin{equation}
m_{k}=\int_{0}^{E_{max}}R^{L}(E^{'})E^{'k}dE^{'}~, \label{eq3}
\end{equation}
E$_{max}$ is the maximum excitation energy, which is carried out
till 60 MeV in the present calculations. From those moments, then,
we can define the centroid energy of the response function,
\begin{equation}
\overline{E}=m_{1}/m_{0}~. \label{eq4}
\end{equation}

 We apply the fully consistent relativistic RPA
to the collective excitations, especially in the isovector GDR
mode. The isovector dipole operator is $Q=\gamma^0rY_{10}\tau_0$,
which excites an $L=1$ type electric (spin-non-flip) ($\Delta T=1$
and $\Delta S=0$) giant resonance with spin and parity $J^\pi
=1^{-}$. The response functions of the nuclear system to the
external operator are calculated at the limit of a small momentum
transfer. It is also necessary to include the space-like parts of
vector mesons in the relativistic RPA calculations, although they
do not play role on the ground state\cite {MGR01}. The
self-consistent treatment guarantees the conservation of the
vector current.

First, we show nuclear matter properties: incompressibility
$K_{\infty}$, effective mass $M^{*}/M$, and symmetry energy
$a_{asy}$ in Table I, which are calculated in the RMF with various
non-linear parameter sets: NL-SH, TM1, NL3, and NL1. The
calculated centroid energies of the isovector GDR in the
relativistic RPA with those non-linear models for some double
closed shell nuclei: $^{208}$Pb, $^{116}$Sn, $^{90}$Zr, and
$^{40}$Ca are also listed in Table I. We also compare the
calculated values with the corresponding experimental data. It is
found that the relativistic RPA with those effective interactions
can to a great extent give a good description of the experimental
results for the centroid energies of the isovector GDR. According
to the macroscopic hydrodynamics models\cite{Gold47} the restoring
force of the giant dipole mode is proportional to the symmetry
energy of nuclear system. Although the nuclear symmetry energies
for the four parameter sets vary from 43.6 MeV to 36.1 MeV, all of
them give more or less reasonable centroid energies of the
isovector GDR  once they could well describe ground state
properties of nuclei. The results of the centroid energy of the
isovector GDR do not show a strong relationship between excited
energies of GDR and the nuclear symmetry energy. Actually, even
the symmetry energy at saturation density is not well constrained
experimentally. Recently, it is reported\cite{Vre03} theoretically
that the nuclear matter symmetry energy at saturation density
(volume asymmetry) can be located in the range of 32 MeV $\leq$
a$_{4}$ $\leq$ 36 MeV based on effective Lagrangians with
density-dependent meson-nucleon vertex function in the framework
of RMF theory. Up to now the density dependence of the nuclear
symmetry energy is still under investigation by various
theoretical approaches\cite{Sat02,Tod03}.

\begin{table}[hbtp]
\begin{tabular}{cccccc}
\hline \hline
 &NL-SH& TM1&NL3&NL1&Exp.\\
\hline
$K_{\infty}[MeV]$&355.36&281.0&271.76&211.29&\\
$M^{*}/M$&0.597&0.634&0.600&0.570&\\
$a_{asy}[MeV]$&36.1&36.9&37.4&43.6&\\
\hline
$^{208}$Pb&13.43&13.07  &13.16 &13.83&13.3$\pm$0.1  \\
$^{116}$Sn&16.04&15.71&15.77&16.05&15.7$\pm$0.2\\
$^{90}$Zr&17.47&17.09&17.19&17.59&16.5$\pm$0.2\\
$^{40}$Ca&19.97&19.79&19.57&19.83&19.8$\pm$0.5\\
\hline
 \hline
\end{tabular}
\caption{Nuclear matter properties: incompressibility
$K_{\infty}$, effective mass $M^{*}/M$, and symmetry energy
$a_{asy}$, calculated in the RMF with various parameter sets. The
centroid energies of the isovector GDR for $^{208}$Pb, $^{116}$Sn,
$^{90}$Zr, and $^{40}$Ca are calculated in the relativistic RPA
with the corresponding parameter set, respectively. The
experimental data are taken from Ref.\cite{Ber75}. All energy
values in the table are in unit of MeV.}
\end{table}

We also apply the relativistic RPA to make a systematic
investigation on isovector dipole mode for stable nuclei along
$\beta$-stability line. It is known that the best results have
been obtained with the NL3 effective interaction for ground state
properties as well as the collective giant resonance. Therefore,
we shall employ the NL3 effective interaction to perform the
relativistic RPA calculations for response functions of isovector
dipole mode. In our investigation we consider even-even, odd and
odd-odd nuclei, which are stable against $\beta$-decay in a wide
range of mass number $12<A<240$. There are more than 130 nuclei
under our investigation, including spherical, deformed, and some
stable isotope nuclei. In our calculations a spherical symmetry is
assumed to simplify the relativistic RPA calculations, because
most of nuclei we are dealing with are of closed shell or
sub-closed shell, even are open shell. For some open-shell nuclei
the last single particle level is partially occupied, the
excitation of a deeper state to this partially occupied state is
allowed. In order to take account of such excitations, an average
assumption is adopted and each wave function is multiplied by its
occupation factor\cite{MTC97}.

\begin{figure}[hbtp]
\includegraphics[scale=0.3]{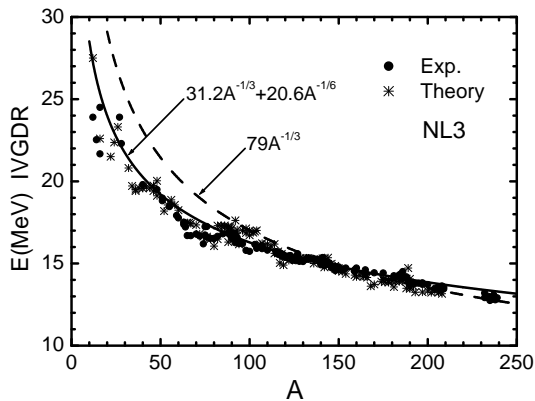}
\vglue -3.0cm \caption{\small{The centroid energies of the
isovector giant dipole resonance for various nuclei along the
$\beta$-stability line as a function of nuclear mass number. The
solid-circles represent the experimental data. The theoretical
results calculated with parameter set NL3 are indicated by stars.
The lines (solid and dash) represent the results of the giant
resonance energy as a function of nuclear mass number calculated
with different systematic behaviors, see the text for details.}}
\label{travagliosm}
\end{figure}

The centroid energies of the isovector GDR for nuclei along
$\beta$-stability line as a function of nuclear mass number are
plotted in Fig.1. The theoretical results calculated with the
parameter set NL3 in this work are indicated by stars. The
solid-circles represent the experimental data, which are taken
from Ref.\cite{Ber75}. It is noted that some heavy nuclei with
mass number located in the region of $150<A<200$ as well as some
light nuclei are deformed nuclei. The response functions of the
isovector GDR for deformed nuclei are split into two peaks  due to
the fact that the vibration frequencies along the long-axis and
the short-axis are different. The experimental centroid energies
given in Fig.1 for those deformed nuclei are the average value of
that two excited energies\cite{Ber75}.  As can be seen from the
figure, the theoretical results calculated with parameter set NL3
agree with the experimental data perfectly  not only for heavy
nuclei but also for medium-weight nuclei. For light nuclei with
mass number is less than 40, the theoretical result is not in a
good agreement with experimental data. In fact, the distributions
of response functions of isovector GDR for those nuclei are very
fragmented and less collectivity, it is difficult to extract the
centroid energies of isovector dipole mode experimentally. This
might be the reason of difference between theoretical and
experimental results for light nuclei.

On the other hand, the peak energy of the isovector GDR has been
predicted by various collective models of giant resonances to be
proportional either to $A^{-1/3}$ or $A^{-1/6}$. However, it is
found experimentally that the excited energy can be well described
by the relationship combined between $A^{-1/3}$ and $A^{-1/6}$,
read as,
\begin{equation}
E_{m}=31.2A^{-1/3}+20.6A^{-1/6}~. \label{eq5}
\end{equation}

In Fig.1, we give the result calculated by eq(5), which is plotted
by a solid curve. At the same time, we also show the curve
calculated by the systematic behavior, $E_{m}=79A^{-1/3}$, which
is failed to describe the experimental data for the medium and
light nuclei as can be seen in the figure.

In a summary, we have studied the isovector GDR for some double
closed shell nuclei: $^{208}$Pb, $^{116}$Sn, $^{90}$Zr, and
$^{40}$Ca in a fully consistent relativistic RPA with different
effective Lagrangians recently developed. By compare the centroid
energies of those nuclei with experimental data, it is found that
the parameter set NL3 can perform a good description on the
properties of isovector GDR. The centroid energies of nuclei along
$\beta$-stability line calculated with the parameter set NL3 are
compared with corresponding experimental data and make a
systematic investigation. The general conclusion is that the
effective Lagrangian with an appropriate nuclear symmetry energy,
which can well describe the ground state properties of nuclei,
could also reproduce the isovector giant dipole resonances of
nuclei along the $\beta$-stability line.

\vglue 0.5cm
 This work is supported by the National Natural Science
Foundation of China under Grant Nos 10075080 and 10275094, and
Major State Basic Research Development Programme in China under
Contract No G2000077400. One of the authors (CAO) wishes to thank
Prof. Zhang Zong-ye and Prof. Yu You-wen for many stimulating
discussions and Prof. Su Zong Di for supplying us with the
experimental data.

\end{document}